\documentclass[fleqn,10pt]{wlscirep}
\usepackage[utf8]{inputenc}
\usepackage[T1]{fontenc}
\usepackage{subcaption}
\usepackage{threeparttable}

\title{User Exploration and Exploitation Behavior Under the Influence of Real-time Interactions in Live Streaming Environments}

\author[1,*]{Akira Matsui}
\author[2]{Kazuki Fujikawa}
\author[2]{Ryo Sasaki}
\author[2]{Ryo Adachi}
\affil[1]{Center for Computational Social Science, Kobe University, Kobe, Hyogo, Japan}
\affil[2]{DeNA Co., Ltd., Tokyo, Tokyo, Japan}
\affil[*]{corresponding author amatsui@rieb.kobe-u.ac.jp}


\begin{abstract}
Live streaming platforms offer a distinctive way for users and content creators to interact with each other through real-time communication. While research on user behavior in online platforms has explored how users discover their favorite content from creators and engage with them, the role of real-time features remains unclear. There are open questions as to what commonalities and differences exist in users' relationships with live streaming platforms compared to traditional on-demand style platforms. To understand this, we employ the concept of Exploration/Exploitation (E/E) and analyze a large-scale dataset from a live streaming platform over two years. Our results indicate that even on live streaming platforms, users exhibit E/E behavior but experience a longer exploration period. We also identify external factors, such as circadian rhythms, that influence E/E dynamics and user loyalty. The presented study emphasizes the importance of balancing E/E in online platform design, especially for live streaming platforms, providing implications that suggest design strategies for platform developers and content creators to facilitate timely engagement and retention.
\end{abstract}

\begin{document}

\flushbottom
\maketitle

\thispagestyle{empty}

\section{Introduction}

Online platforms have deeply permeated modern society, affecting everyday life and cultural trends. Not only do they provide a remarkable range of content, from text to movies, but they also serve as environments that allow users to interact with digital content, creators, and other users. In particular, user-generated content (UGC) platforms, such as Twitter~\cite{hu2012breaking}, Reddit~\cite{hamilton2017loyalty}, and YouTube~\cite{ma2021advertiser,ma2021advertiser,wu2019agent}, allow users not only to consume content but also to act as content creators. This interactive functionality of UGC platforms enhances users’ ``loyalty'' to content creators or the platforms themselves, which is defined as a user's strong and sustained preference for a community or creator over others~\cite{hamilton2017loyalty, karnstedt2010churn, dror2012churn, danescu2013no, arguello2006talk, backstrom2013characterizing}. This growing body of research leverages these ecosystems to examine how users discover and engage with content, as well as how their patterns of interaction influence subsequent behavior.

While the literature quantitative analysis of user behavior on online platforms contributed to platform design, such as computer-human interaction or user experience (UX), previous studies have focused mainly on platforms that provide content in an on-demand style. On-demand style platforms, like Netflix and Spotify, serve content to users anytime at their convenience. However, in recent years, forms of online content consumption have diversified, and live-streaming platforms have witnessed substantial growth~\cite{marketresearchfuture_2023} and received increasing attention from HCI research community~\cite{mallari2021understanding,wang2019love,lessel2017expanding,hamilton2014streaming,tang2016meerkat,striner2021mapping,lu2018you,lu2019feel} because it provides a user experience different from traditional on-demand style platforms.

The crucial aspect of live-streaming platforms is that users consume content under the real-time interaction constraint, ``watching what is currently being streamed'' while simultaneously engaging in two-way communication with both the streamer and other viewers. \cite{torhonen2021streamers} characterize this as ``real-time interactions,'' highlighting how it enables real-time interaction between streamers and viewers. Furthermore, this real-time social interaction alters communications between creators (live streamers) and content consumers (viewers) from passive to active~\cite{smith2013live, tang2016meerkat} and strengthens their connection~\cite{wulf2020watching,wohn2020live, chen2018drives} through functions like real-time chat or gifting~\cite{hamilton2014streaming,tang2016meerkat}. This characteristic of real-time interaction potentially acts as a parameter in user behavior, altering user models previously proposed based on findings in on-demand platform environments. For example, real-time interaction may further increase loyalty to creators due to its dense, simultaneous communication. Moreover, since users on live-streaming platforms can only consume content being streamed live, they might take more time to nurture their attachment to creators than on-demand platforms.

Nonetheless, many live-streaming sites currently rely on UX designs and recommendation algorithms mainly developed for on-demand platforms. This implies that the design of live-streaming platforms assumes that user behavior and human-computer interaction are largely the same as those observed in on-demand style platforms, but it remains unclear in which aspects the two differ and align. In other words, much of the human-computer interaction on streaming platforms is based on knowledge from on-demand platforms, and its validity in this new context is uncertain.

To fill this knowledge gap, this study aims to investigate how users develop their relationships with live-streaming content and creators through the lens of exploration and exploitation (E/E). E/E are well-established concepts first introduced in organizational studies~\cite{march1991exploration,mom2007investigating} and applied in online platforms~\cite{GOMEZZARA2024108014}. Exploration involves seeking new content, creators, or ideas that align with their interests, while exploitation focuses on revisiting and deepening engagement with content they already enjoy. This balance enhances users' experiences and strengthens their ties to both the platform and its creators. The concept of E/E allows us to capture complex user behavior in a simple yet formal way, and it is widely utilized in computer science literature, such as recommendation systems for domains like e-commerce~\cite{wang2019modeling}, movie streaming~\cite{pereira2022efficient,broden2019bandit}, web crawling~\cite{schulam2023improving}, music listening~\cite{mok2022dynamics,anderson2020algorithmic}, social media~\cite{GOMEZZARA2024108014}, and short videos~\cite{suLongTermValue2024}. While this body of literature focuses on on-demand platforms, we know little about the dynamics of E/E on live-streaming platforms and whether they differ. Users on live-streaming platforms face the real-time interaction constraint, which can limit content availability for exploration or exploitation of creators. This raises questions about whether E/E behavior occurs on live-streaming platforms and whether it mirrors the dynamics observed on on-demand platforms. For example, compared to on-demand platforms, the pace of exploration on live-streaming platforms may be slower because users can only explore the content that is live (the real-time interaction constraint). In addition, since the amount of available content fluctuates throughout the day, the supply of live content affects their E/E behavior, which users are unable to control. With these notions, our study will investigate whether factors specific to live-streaming content can affect users' E/E behavior and how this is linked to their formation of loyalty to content creators (live streamers).

By harnessing the two well-established concepts of user behavior in online platforms—E/E and loyalty—we study the following research questions:
\begin{itemize}
    \item RQ1: How do users on live-streaming platforms adjust their balance of exploration and exploitation under the conditions of real-time content delivery?
    \item RQ2: How do changes in exploration and exploitation behaviors affect the process of forming loyalty toward creators and the platform?
\end{itemize}

By answering these research questions, we not only aim to provide empirical evidence of user behaviors in live-streaming contexts but also generate actionable insights to improve UX designs and functions, such as recommendation algorithms, that enhance engagement through real-time interactions.

\vskip\baselineskip
\noindent
\textbf{The Present Work} This study aims to uncover the mechanisms behind users' digital consumption on a live-streaming platform through the lens of E/E behavior. We explore a large dataset from a live-streaming platform, comprising two years of longitudinal user history (over 584M live-streaming histories of 2.2M users). Specifically, our goal is to comprehend the process by which users who engage with live streams choose streamers and cultivate their loyalty toward live streamers. We investigate the dynamics of E/E behavior on the live-streaming platform and compare the results with an on-demand style platform (a music listening platform). To capture user-level E/E behavior dynamics, we calculate the turnover rate, which is the ratio of the consumption of new content to already consumed content~\cite{mok2022dynamics}. Furthermore, we analyze the streamers' career trajectories as they attract users who are exploring live-streaming content. The present analysis also illuminates an external factor—the balance between demand (users watching live streams) and supply (live streamers providing streaming videos)—that can affect users' E/E behavior represented by the turnover rate.

\vskip\baselineskip
\noindent
\textbf{Overview of The Results}
Our investigations into live-streaming platforms provide critical insights into human behavior on online platforms. The findings to be presented illustrate a consistent balance of exploration and exploitation convergence as users stay on the platform. This pattern highlights a ``golden period'' of users' exploration, during which streamers have a heightened opportunity to cultivate a dedicated audience. Moreover, our analysis suggests that, in addition to individual preferences, external factors can affect users' E/E behavior. Additionally, we identify a link between user loyalty towards live streamers and their E/E behavior. Overall, this research offers valuable insights for both the academic literature and administrators of live-streaming platforms

\vskip\baselineskip
\noindent
\textbf{Terminologies: }
To make our paper easy to follow, we would like to introduce several terminologies regarding the platform of interest. We summarized them in Table~\ref{tab:table1}. In the context of live streaming platforms, a notable complexity arises from the presence of two distinct categories of participants: those who engage in streaming live content (referred to as ``live streamers'') and those who consume this content (hereinafter termed ``users''). In addition, we define exploration and exploitation in this study.

\begin{table}[ht]
\centering
\begin{tabular}{p{5cm}p{8cm}}
\textbf{Term} & \textbf{Definition} \\ \hline
Users & Individuals on the platform who primarily watch live streams. \\
User age & The age of a user that starts the day the user joins the platform. \\
Live streamers & Participants on the platform who broadcast live content. \\
{User \(i\)'s recurring live streamer \(k\)} & {Live streamer \(k\) is a recurring live streamer for user \(i\) when user \(i\) has watched at least one of \(k\)'s live sessions, indicating user \(i\)'s consumption of \(k\)'s content.} \\
{User \(i\)'s new live streamer \(j\)} &
{A live streamer \(j\) if user \(i\) watches that streamer's live for the first time.} \\
{Exploration} & {The process of searching for and viewing live streaming sessions from user's new live streamers.} \\
{Exploitation} & {The process of viewing live streaming sessions from a user's recurring live streamers.} \\
{The real-time interaction constraint} & {The constraint that users on live streaming platforms can only watch sessions while they are live.}\\

Highest rank of live streamer \(j\)&
The highest rank live streamer \(j\) ever achieved in that live streamer's career.
\end{tabular}
\caption{Terminologies of this Research}
\label{tab:table1}
\end{table}

\section{Background}

In this section, we discuss two pivotal aspects of user behavior research for the present study: E/E behavior and user loyalty on online platforms. These concepts are essential for understanding how users engage with diverse forms of content across various platforms, from e-commerce to social media and on-demand content services. The following subsections describe these behaviors and their implications for the design and success of digital platforms.

\subsection{Exploration and Exploitation (E/E) Behavior}

We first review E/E behavior, primarily in the context of user behavior on digital platforms. As mentioned in the introduction, the E/E concept, originating from organizational studies~\cite{march1991exploration,mom2007investigating}, has been applied across a wide range of disciplines, from recommender systems~\cite{wang2019modeling, pereira2022efficient,broden2019bandit,anderson2020algorithmic} to studies on creativity~\cite{liu2021understanding}.

The concept of E/E models consumption behavior by dividing it into two phases: exploration and exploitation. During the exploration phase, users search for new items to consume. For instance, when you explore new music tracks on a music streaming platform, you are in the exploration phase. Conversely, when listening to music you have already listened to, you are in the exploitation phase. If users change their preferences regarding consumed content, they may transition from exploitation phases. \cite{GOMEZZARA2024108014} describe E/E behavior in terms of uncertainty: exploration involves seeking new content at the expense of uncertainty, while exploitation involves consuming known content in search of certainty. The concept of E/E has a wide range of applications. \cite{wang2019modeling}, for example, propose a model capturing the recurrent nature of consumer behavior, while \cite{benson2016modeling} focus on the pattern of declining repeated consumption of digital content over time.

The E/E behavior also explains the process by which users find digital content and become attached to it, such as in music listening~\cite{anderson2021just, ferwerda2017personality}. Understanding the mechanisms behind repeated consumption and its fluctuations has both theoretical and practical implications for research on human behavior in digital content consumption. A notable recent example is \cite{mok2022dynamics}, which study E/E behavior in music listening by calculating the turnover rate, a method also utilized in this paper.

Longitudinal research on user consumption demonstrates that preferences or tastes can evolve over time. For example, \cite{mcauley2013amateurs} show how tastes evolve, highlighting the need to consider a user's evolving experience and expertise in product recommendations. \cite{moore2013taste} propose a dynamical model that analyzes and visualizes the evolution of listening preferences using eight years of Last.fm data and uncovers trends in user tastes. Similarly, \cite{schedl2015tailoring} and \cite{holtz2020engagement}
investigates the evolution of user preference for music listening.

The literature suggests that mixing exploration and exploitation improves the performance of music recommendation~\cite{pereira2022efficient,broden2019bandit} and web crawling~\cite{schulam2023improving}. This principle is also supported theoretically~\cite{immorlica2019diversity}. Therefore, achieving a balance between exploration and exploitation is essential. In this vein, \cite{GOMEZZARA2024108014} conduct an empirical study on Snapchat users, combining survey data and user logs to understand the interplay between these behaviors, emphasizing the need to consider exploration and exploitation separately. This study reveals that the degree of E/E observed changes within a day. The concept of E/E is also crucial for UGC platforms, as users engage in exploration and exploitation not only in content consumption but also in content creation~\cite{xiang2022user}.

\subsection{Loyalty to Digital Platform}
Understanding user loyalty is a central issue for digital platforms as it is crucial for retaining new users (i.e., reducing churn)~\cite{karnstedt2010churn, dror2012churn, danescu2013no} and enhancing user engagement~\cite{arguello2006talk, backstrom2013characterizing}. For instance, \cite{hamilton2017loyalty} explores loyalty within Reddit communities, demonstrating that loyal users and communities exhibit unique behaviors and interaction patterns. In the context of the creator economy, \cite{el2022quantifying} study the mechanism of the membership platform of the creator economy including loyalty and retention. The recent literature suggests that creators develop priorities among platforms~\cite{ma2023multi}, and therefore studying the loyalty of users to creators has been an important issue in the literature and 

Much of the research on live streaming platforms employs psychological theories to understand user motivation~\cite{lu2018you}. For example, \cite{lim2020role} utilize cognitive theory and parasocial relationship models to uncover that users' identification and emotional engagement significantly impact their interaction with live streaming. Additionally, research in e-commerce live streaming has formed a substantial segment, given that user loyalty is critical for business success in this domain~\cite{wongkitrungrueng2020role, li2021attachment}. From a community perspective, \cite{hilvert2018social} identifies six motivations driving live-stream viewer engagement on Twitch, underscoring the social and community aspects of these motivations, especially among viewers of smaller channels.

\subsection{ Real-Time Interaction in Community Building on Live Streaming Platforms}
Our work bridges several lines of literature discussed in this section. 
Live streaming platforms offer a unique environment where users actively engage with real-time social interaction~\cite{smith2013live, tang2016meerkat}, which strengthens the connection between streamers and viewers~\cite{wulf2020watching,wohn2020live, chen2018drives}. Functions powered by real-time interactions like real-time chat or gifting play a pivotal role in user engagements in live streaming platforms~\cite{hamilton2014streaming,tang2016meerkat}. Particularly, the literature has pointed out that this interactive communication induces the emergence of temporary and ad hoc communities~\cite {haimson2017makes,hamilton2016rivulet,hamilton2014streaming,tang2016meerkat}. For instance, \cite{hamilton2014streaming} outline that real-time interaction in live streaming helps build community by allowing streamers and viewers to connect instantly and share experiences. 

\subsection{Relation to This Work}

While real-time interaction builds unique and highly intensive communication environments, it also imposes a strict limitation that they can only watch live streams being lived (the real-time interaction constraint). This distinctive characteristic may affect users' content consumption behavior. To understand this, we study the E/E behavior, which the expanded body of research emphasizes its importance and highlights it may vary according to the type of platform. In addition, the literature suggest live streamers leverage the real-time interactions to form communities and enhance the relationships with users. This raises a question that how users form their loyalty to live streamers (content creators) under the real-time communications and research on user loyalty in online platforms predominantly study on-demand style platforms such as Twitter/X or Reddit. 

\section{Data and Methods}\label{sec:data}
This section describes the data and methods used in this study. We first introduce the live streaming platform from which we obtained the data and explain the relevant terminologies. We then discuss the metrics and statistical models used for our analysis.

\subsection{Dataset of the Live Streaming Platform}
\begin{figure}[ht]
  \centering
  \includegraphics[width=0.6\columnwidth]{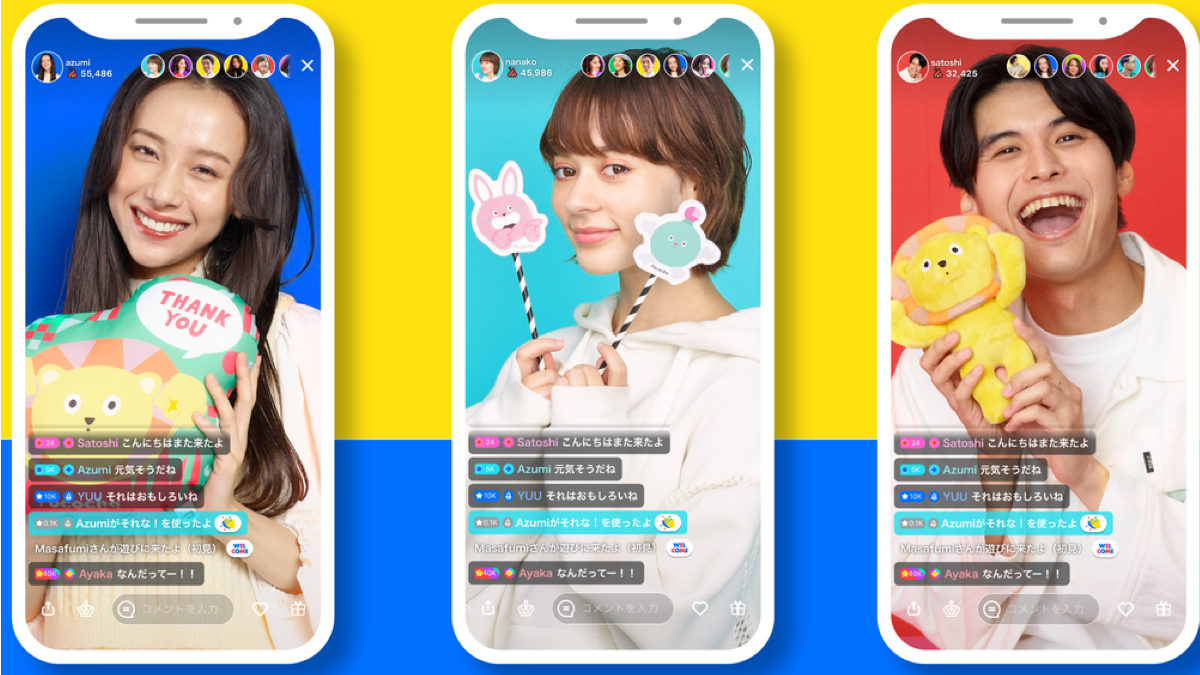}
  \caption{Examples of live streaming session}
  \label{fig:figure1}
 
\end{figure}

In this study, we study the data from Pococha, a digital platform for online live streaming amassing over 5.34 million total app downloads as of September 30, 2023. In the platform, live streamers can interact with users through chat or purchased items (monetary reward) during their live stream. Users can follow live streamers and re-visit when their favorite live streamers broadcast again. This allows us to study live streaming behavior, where the same live streamers consistently go live, and the same users regularly watch these live streaming sessions. In other words, this platform hosts multiple live-streaming sessions by the same live streamer, and the same users repeatedly watch them. The platform is accessible exclusively through smartphones, prohibiting usage from PCs or other devices. Consequently, both broadcasting and viewing of live streams are conducted solely via smartphones or tablets (iOS or Android OS). Hence, all user behaviors analyzed in this study are those undertaken on smartphones or tablets. The platform predominantly operates in Japan, the United States (closed on February 2024), and India. These three areas are separated on the platform and this study focuses on the Japan-based area.

Thanks to our concentration in a single country (Japan), we ensure that all users in the dataset use the same language, timezone, and currency. While our data comes from a country that shares a relatively homogeneous cultural context compared to some other more diversified ones such as the US, we acknowledge the importance of accounting for heterogeneity in the data to ensure robust and generalized results. To address this, we employ a rigorous statistical model described in Section~\ref{sec:fixed_effect_method} to control for potential confounding factors. 

We would like to note here that we obtained and analyzed the data from the platform within the users' consent, following the privacy policy of the data provider~\cite{denaPrivacy} and the Japanese law~\cite{denaPrivacyJP}. The dataset contains users' live streaming watch history and live streamers' live history. In addition, it contains the live streamers' ranking history that is to be used and detailed in Section~\ref{sec:streamer_ranking}. The study was exempt from ethical review by the Institutional Review Board, in accordance with the author institutional guidelines, as it did not involve human subjects. Our data set contains the users who use iOS or Android OS in the Japan-based platform. As a result, our data set consists of over 584M live streaming histories from over 2.2M Japan-based users, spanning 2 years between 2020 JAN 1 and 2021 DEC 31. 

\subsection{Music Listening Dataset}
This study also analyzes an on-demand style platform to understand the E/E behavior without real-time interactions. For this purpose, we utilize the Music Listening Histories Dataset (MLHD)~\cite{vigliensoni17music}, which comprises music listening histories constructed from timestamped logs extracted from Last.fm under the Last.fm API license. In this paper, we analyzed over 24.4 billion data points with over 526k users during the year 2006 to 2012.

\subsection{Turnover Rate and E/E Behavior}\label{sec:tr_ee_method}

To model users' exploration/exploitation (E/E) behavior, we focus on the balance between exploration and exploitation. In this paper, we consider the exploration behavior as watching the live streaming session by the live streamers that user \(i\) has never watched. Conversely, we consider the exploitation behavior as watching the live streaming session by the live streamers whose sessions user \(i\) has previously watched.

To quantitatively capture the balance between exploration and exploitation, we calculate the turnover rate \(TR_{i,w}(t)\) for a time window \(w\) at time \(t\), where \(w\) and \(t\) share the same time interval. The turnover rate \(TR_{i,w}(t) \equiv s_{i,w}(t)/(s_{i,w}(t) + r_{i,w}(t))\) is defined for user \(i\), where \(r_{i,w}(t)\) is the number of the user $i$'s recurring live streaming (exploitation) at time \(t\) during window \(w\), and
\(s_{i,w}(t)\) is the number of live sessions of a streamer that user \(i\) attends for the first time ever in their user life  (exploration) within the time window \(w\) and at time \(t\).

\(TR_{i,w}(t)\) can take a larger value than \(TR_{i,w}(s)\), where \(s<t\). For example, during time window \(w\) (week \(t\)), if user \(i\) watches live streaming sessions by 6 new live streamers whose sessions user \(i\) has never watched before and also watches live streaming sessions by live streamers whose sessions user \(i\) has watched 9 times, then \(s_{i,w}(t)\) is 6 and \(r_{i,w}(t)\) is 9, so \(TR_{i,w}(t)\) becomes 0.4. In the next time window \(t+1\), \(s_{i,w}(t)\) can be 9 and \(r_{i,w}(t)\) can be 6, resulting in \(TR_{i,w}(t) = 0.6\), since these metrics are calculated on a time window basis.

We set the time window \(w\) based on the time unit for the analysis. For the analysis in Section~\ref{sec:a_ee}, 

we set the time window to 1 week, while in Section~\ref{sec:res_oshi}, we set it to 1 month since the metrics used in that section are based on monthly data. Additionally, we calculate a macro-level turnover rate in Section~\ref{sec:tr_cyrcadian}.

It is important to note that during the period of the dataset, the platform did not utilize turnover rate in its functionalities or operations, including recommendations. However, some platform-level factors can alter user behavior, and they are not only at the platform level such as seasonality. To consider and assess these confounding factors, the presented analysis first describes the turnover rate over the course of the dataset period to understand if the observed dynamics of the turnover rate is a stable phenomenon and conducts the same analysis with the open dataset to understand the generality of the detected dynamics in Section~\ref{sec:convergent_tr}. In the user-level analysis, we employ a statistical model that considers user-level differences, such as differences in preferences among users. We use linear regression models with fixed effects, allowing us to incorporate not only user-level confounding factors but also platform-level ones such as seasonality, and we will detail the specification in Section~\ref{sec:fixed_effect_method}.

\subsection{Career Trajectory of Live Streamers and Users' Loyalty to Live Streamers}

We are also interested in the relationship of E/E behavior to other important aspects: creators' performance and users' loyalty to the creators. To this end, we conduct two studies: one on live streamers' career trajectories and the other on users' loyalty.

\subsubsection{Career trajectory of Live Streamers}
The first study focuses on the performance of live streamers over the course of their career trajectory and the degree of user turnover. We utilize the live streamer rank determined by the platform's monthly rankings, based on their achievements in live streaming, such as the amount of support (items) they obtained and the number of users they attracted within a month.

Live streamers are ranked in the order of E, D, C, B, A, and S, with S being the highest rating. The live streamer rank serves as an ascending indicator of live streamers' monthly performance. In our analysis, we will focus on the live streamers who ever achieved S, A or B. 
Focusing on these live streamers allows us to understand the career trajectories of top live streamers and their performance.

\subsubsection{Users' Loyalty to Creator (Oshi)}

We also consider the relationship between the E/E behavior and users' loyalty to the creators (i.e., live streamers). While users' loyalty to platforms has been an important issue in computational social science~\cite{hamilton2017loyalty}, we focus on loyalty to creators as an analogy of the creator economy~\cite{el2022quantifying}. On this platform, both monetary and intangible rewards, such as user comments, are pivotal. Users can select their favorite live streamer and reward them monthly. They can also support multiple streamers, to whom they are strongly committed. 

A live streamer who receives such dedicated support is called an ``Oshi,'' a term rooted in pop culture and prevalent in fandom communities~\cite{UrbanDictionary2017} such as idle culture~\cite{yakura2021no}.\footnote{ Academically, ``Oshi'' derives from the Japanese verb ``osu,'' meaning ``to push.''} The platform has criteria to designate a live streamer as a user's Oshi, detailed in Section~\ref{sec:criteria_oshi}. Understanding how users designate their Oshi based on turnover rate can reveal the relationship between their loyalty to creators and the E/E behavior. To measure this loyalty, we calculate the number of live streamers they consistently support.

\subsection{Fixed Effect Linear Regression}\label{sec:fixed_effect_method}
{
This subsection explains the fixed-effect linear regression models, which are well-developed in the econometric literature~\cite{hanck2019introduction,hansen2022econometrics}. More precisely, the analysis conducted in this study with the regression model is known as panel data analysis~\cite{hansen2022econometrics}.
We employ this method to understand the dynamics of exploration and exploitation while controlling for potential confounding factors that may introduce biases in our results. Our panel analysis incorporates user-specific factors (users' fixed effects), such as gender and preferences, and time-specific factors (time-fixed effects), such as events. The model with the user and time-fixed effects mitigates their impact on the dependent variables. The regression is modeled as follows:

\begin{equation}\label{reg:page-creation}
y_{\mathrm{it}} = \sum_{\substack{s}} \gamma_s x_{\mathrm{its}}
+ \mu_i + \mu_t + u_{\mathrm{it}}
\end{equation}
\noindent
Here, \( x_{its} \) represents the vector of independent variables \( s \) for user \( i \) at time \( t \), with \( \gamma_s \) as parameters of interest; \( \mu_i \) is the fixed effect specific to User \( i \), and \( \mu_t \) is the fixed effect for time \( t \); \( u_{it} \) is the error term; \( y \) is the dependent variable.

In this paper, we introduce two fixed-effect models. We employ the first model Section~\ref{sec:tr_oshi} and examine the factors associated with changes in Oshi. It is natural to consider users' decisions to select live streamers for Oshi; not only turnover rate but also individual preferences or platform-level events matter. This first model incorporates both users' fixed effects and the fixed time effects of each month. The second model estimates the turnover rate over the course of the day but addressing the potential bias in turnover rate due to individual differences, as discussed in Section~\ref{sec:tr_cyrcadian}.
}

\begin{figure*}[t]
  \centering
  \begin{minipage}{\textwidth}
    \centering
    \includegraphics[width=0.99\linewidth]{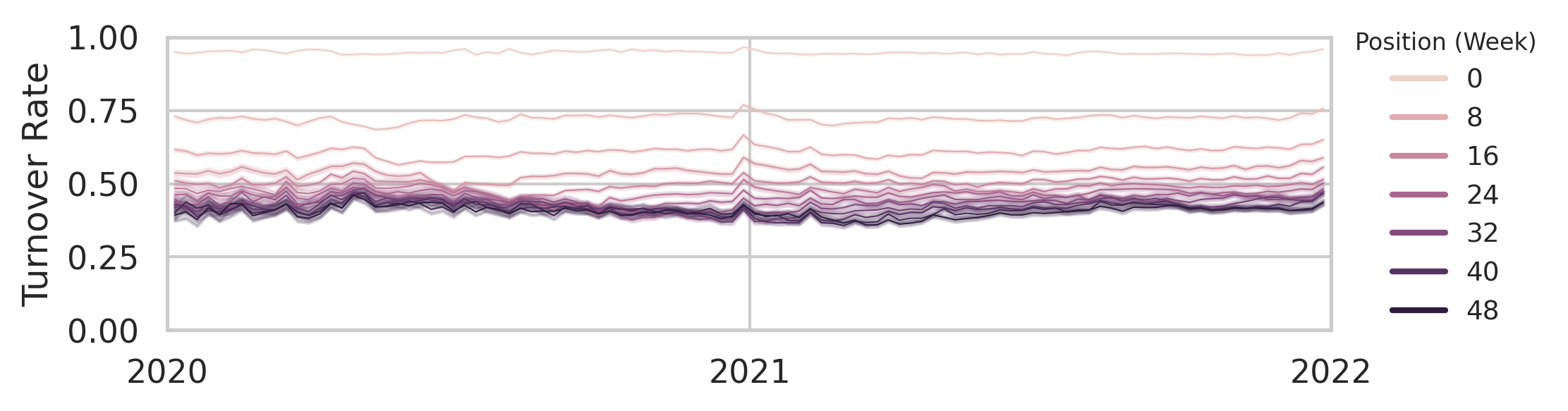}
    \caption{Transition of Turnover Rate: Live Streaming. The transition of average turnover rate per user age on the live streaming. The user age is calculated as the number of weeks after the users join the platform. The color represents the age, and the shadows represent 95\% CIs.}
    \label{fig:figure2}
 
  \end{minipage}
  \vspace{0.1cm}
  \begin{minipage}{\textwidth}
    \centering
    \includegraphics[width=0.99\linewidth]{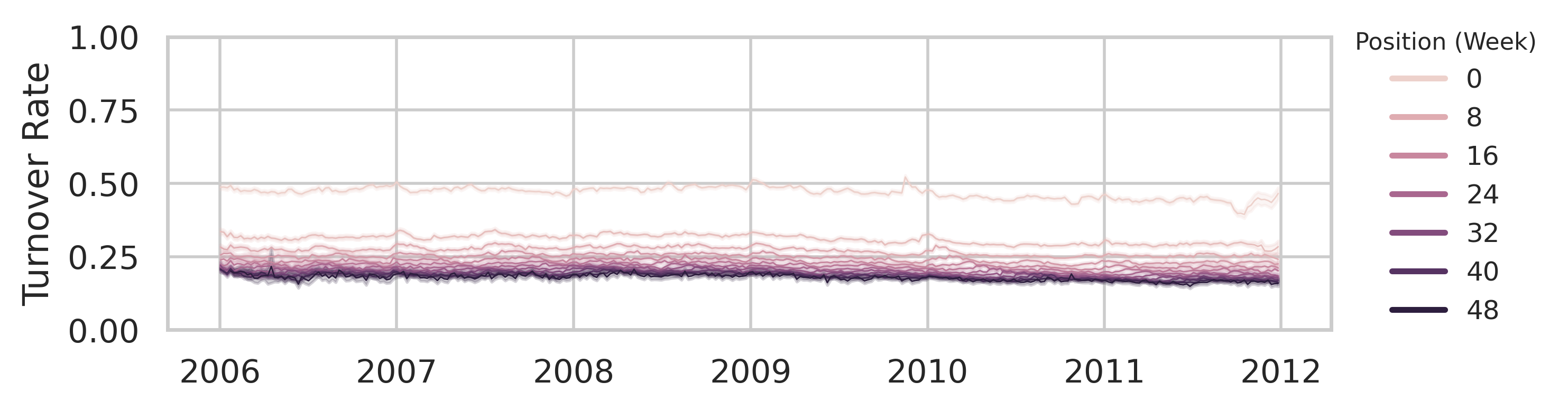}
    \caption{Transition of Turnover Rate: Music Listening. The transition of average turnover rate per user age on the music listening platform (MLHD last.fm dataset). The user age is calculated as the number of weeks after the users join the platform. The color represents the age, and the shadows represent 95\% CIs.}
    \label{fig:figure3}
 
  \end{minipage}
\end{figure*}

\section{Analysis on E/E Behavior}\label{sec:a_ee}
In this section, we discuss the results of our analysis of the E/E behavior as represented by the turnover rate. We first study the dynamics of the turnover rate over the course of the data periods and its intraday dynamics. Then, we examine the career trajectory of live streamers.

\subsection{Convergence of Exploration Activity}\label{sec:convergent_tr}

We investigate the balance between exploration and exploitation behavior throughout users' lifespans on the platform. Figure~\ref{fig:figure2} depicts the average turnover rate decreases as user age increases, signifying a reduction in exploration activity. Within the first two months, users' content consumption predominantly (over 70\%) consists of content from new live streamers.{ This exploration behavior changes as the users stay on the platform. The figure describes that the users eventually stabilize at a 40\% turnover rate. This indicates that 40\% of the live streaming content they engage with originates from new live streamers, while the remaining 60\% involves sessions with recurring live streamers.} Moreover, despite some fluctuations, we observe a consistent association between user age and their exploration activities across the period. We note that the decrease in the turnover rate during the initial several weeks could be due to the cold start problem, where users are not yet familiar with the functions or system of the platform. However, the detected convergence process typically takes about 24 weeks on average (i.e., 6 months). Therefore, the cold start problems concerning users' initial periods do not explain everything.

These findings suggest that the initial weeks post-registration represent a ``golden period'' for both live streamers and platform operators. During this stage, users demonstrate significant exploration behavior, engaging with content from various live streamers while showing minimal commitment to particular ones. This phase is pivotal for streamers as it offers the chance to convert new users into dedicated fans. This period can also provide an opportunity for the platform to improve user retention rates. The identified trend also suggests that long-standing users are less likely to significantly alter their E/E behavior. For instance, strategies aimed at modifying the turnover rate may prove to be less effective for users who have been on the platform for an extended period.

The above finding provides an insightful understanding of E/E behavior on live streaming platforms. First, it is surprising that users on live streaming platforms engage in E/E behavior. Despite the limited content options imposed by the real-time interaction constraint, they manage to maintain a balance between exploring new content and exploiting familiar content afterward. However, it also raises a natural question: {\it can we observe the detected conversion pattern of E/E behavior (turnover rate) on other platforms without real-time interactions?} To answer this question, we analyze the music listening dataset (MLHD) and calculate the turnover rate where $s_{t,w}(t)$ represents the number of music tracks that user $i$ listens to for the first time, and $r_{t,w}(t)$ represents those that user $i$ has already listened to. We plot the trajectory of the turnover rate in Figure~\ref{fig:figure3}.  

The figure depicts similar phenomena to our findings on the streaming platform but with two key differences. First, the turnover rate on the streaming platform converges more slowly than on the music-listening platform. Second, we find that the stable turnover rate is around 0.25 on the music-listening platform, which is lower than that on the live-streaming platform. Although users on both platforms eventually reach a stable balance between exploration and exploitation (convergence), this balance is influenced by the nature of the platforms. While real-time interaction does not account for all observed differences, we believe this analysis underscores its role in shaping users' consumption behavior on online platforms.

\subsection{E/E behavior and User Types}\label{sec:ee_user_type}
While we have investigated the E/E behavior using the turnover rate, we are also interested in user-level analysis, such as finding out which users prefer exploitation or exploration. To do this, we compare the turnover rate by user types, defined in two ways: live streaming experience and active time. On the platform, users who watch live-streaming sessions can also host their own sessions, meaning they can be both consumers and creators of live-streaming content. We study the E/E behavior of users who have hosted a live session. Additionally, we classify users based on their active time, which is determined by the number of live sessions they consume. For each user, we calculate the time range during which they actively watch live sessions and classify this time range as follows: Morning (6 am-12 pm), Afternoon (12 pm-6 pm), Evening (6 pm-9 pm), and Night (9 pm-12 am). Since the active time can reflect users' lifestyles, we are interested in its relation to E/E behavior.

\begin{figure}[htbp]
  \centering
  \includegraphics[width=0.9\columnwidth]{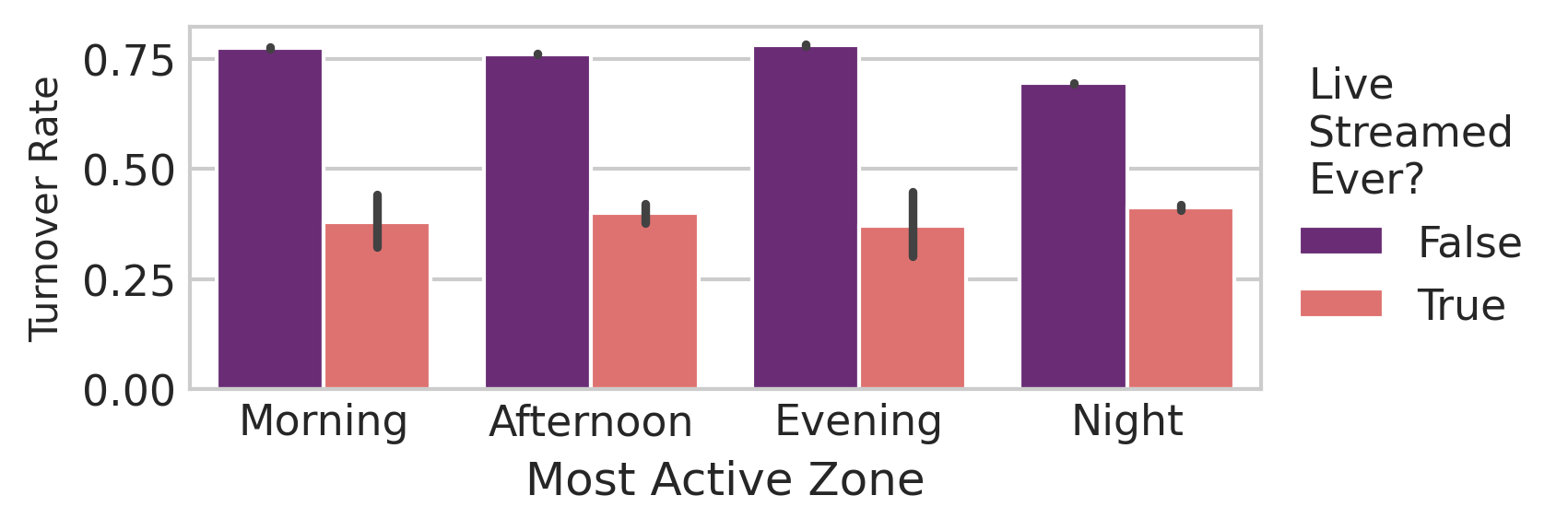}
  \caption{Turnover Rate and User Activities}
  \label{fig:figure4}
 
\end{figure}

We plot the average turnover rate for each user type in Figure~\ref{fig:figure4}. The notable distinction in the figure is that users who have hosted a live streaming session demonstrate a lower turnover rate (around 0.3), indicating they tend to exploit content rather than explore. Users without live streaming experience show nearly average turnover rates across the whole population. These results suggest that users who are also creators tend to exploit others' content and learn from it for their own content creation.

{ This result implies that the degree of exploration varies among user types, but the active time zone is not well linked to the difference in the turnover rate. Users in the same category demonstrate mostly the same turnover rate (i.e., the balance between exploration and exploitation) irrespective of the active time zone, except for users without streaming experience whose active zone is Night. Since the active time zone reflects users' lifestyle or context of watching live streaming, this finding is notable as it suggests that E/E behavior operates independently of these factors.}

We find that users whose active time is at Night show a lower turnover rate, but we do not observe distinctive differences in active time types. However, this result does not imply independence between E/E behavior and time of day because the active time is determined by users' most active time range, meaning they can consume live streaming content regardless of their active time. Since the presented turnover rate is a user-level average, it includes consumption that occurs outside their active time. To clarify the relationship between time of day and E/E behavior, we will conduct a case study in the next subsection.

\subsection{E/E Behavior and External Factors: The Case of Circadian Rhythm}\label{sec:tr_cyrcadian}

\begin{figure*}[tbp]
  \centering
    \includegraphics[width=1\linewidth]{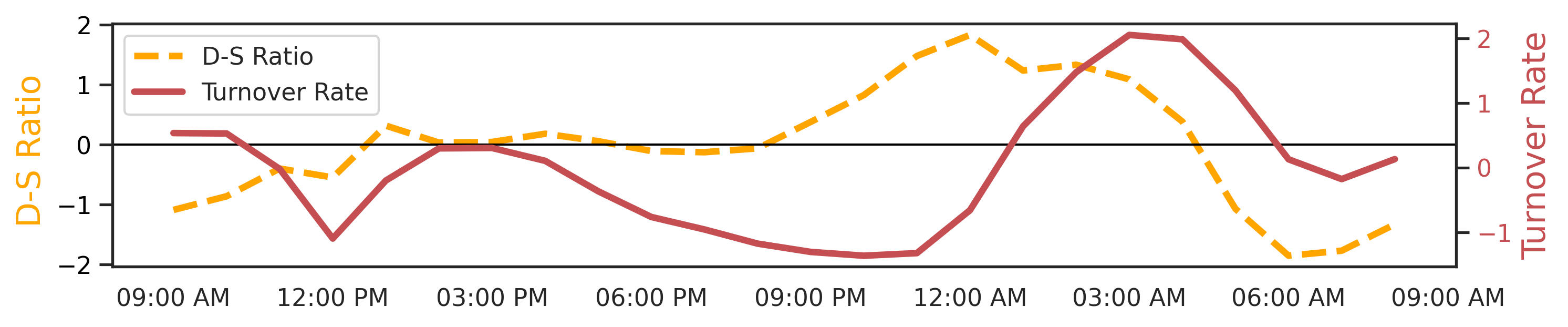}
    \caption{Demand-Supply Ratio and Turnover Rate. The turnover rate and the live streamer-user ratio over the course of the day (z-score). }
    \label{fig:figure5}
\end{figure*}

We examined E/E behavior from the perspective of user lifespan. Our findings suggested that exploration activity tends to stabilize at a certain level as users age on the platform. While such user factors can describe the mechanism behind E/E behavior, external factors are likely to play a pivotal role in this complex behavior. Indeed, the previous subsection led us to study if users change their E/E behavior depending on the time of the day.

Although such within-user factors can describe the mechanism behind E/E behavior, external factors are likely to play a pivotal role in this complex behavior. The recent literature suggests that the circadian rhythm can drive human behavior on online platforms such as content consumption~\cite{piccardi2023curious} or social media use~\cite{zhou2023circadian}. In addition, \cite{GOMEZZARA2024108014} identify the relationship between the circadian rhythm and the E/E behavior of social media users. While they investigate how users on social media consume content, it is not clear whether the constraint that users can only consume content when it is live can explain the fluctuation of E/E behavior in the circadian rhythm. To understand this point, we need to conduct a platform-level analysis that studies the synchronization between the supply-demand of content and the users' E/E behavior. Here, we identify the relationship between the circadian rhythm and E/E behavior. To this end, we will observe that, on a daily basis, the platform experiences fluctuations in users' exploration behavior.

We depict the turnover rate over the course of the day. Here 
we calculate the number of first live sessions for each user \(i\) as \(s_{ia}\) and the number of recurring live sessions for each user \(i\) as \(r_{ia}\) within an interval $a$ where $a$ represents the interval of a day divided into 24 hourly segments (e.g., 1 pm - 2 pm). We then compute the turnover rate as \(TR_a = S_a/(R_{a} + S_{a})\), where \(S_a = \sum_i s_{ia}\) and \(R_a = \sum_i r_{ia}(t)\). We plot the results in Figure~\ref{fig:figure5} and observe a 10\% point fluctuation in the turnover rate from 9 PM to 3 AM. To comprehend the source of the fluctuations, we also plot the number of active users (live watchers) and live streamers that reflect the demand and supply dynamics of the content. In particular, we note that the demand-to-supply (D-S) ratio (users to live streamers) correlates with the turnover rate with lags. The cross-correlation between the two yields a result of 0.914 with a lag of 3, suggesting a change in the D-S ratio can be an early sign of a change in E/E behavior. For example, the figure indicates that the D-S ratio peaks around 12 AM, while the turnover rate peaks around 3 AM. This observation implies that the D-S ratio is a late indicator of the E/E behavior represented by the turnover rate, implying that platforms can use this macro indicator to predict changes in user behavior. We also confirm that this fluctuation is not caused by the alternation of the composition of the users. We find the fluctuations still remain even in the estimated turnover rate with the covariance, and report it in Figure~\ref{fig:figure12}.

This observation has implications for both the literature and practitioners of online platforms. These fluctuations can influence live streamers' decisions regarding their content streaming schedules. Given that users' choices in live streaming content consumption are affected by time-linked fluctuations, streamers might strategically choose their streaming times. Similarly, live streamers who lack flexibility in scheduling may not be able to align their behavior with this macro-level condition. 

The observed correlation also suggests that E/E behavior, represented by turnover rate, can be affected by the number of live streamers active per user on the platform at that time (demand-supply). The 3-hour lag indicates that the impact of demand and supply on the user does not materialize immediately but exhibits stickiness, meaning users' behavior does not instantly react to the platform's state. This observation implies that it might be difficult for platform administrators to manipulate or facilitate E/E behavior in real-time.

While it would be challenging to manipulate E/E behavior through the D-S ratio over time, platforms may use it to understand how users interact with content. The D-S ratio can serve as an early indicator of changes in E/E behavior. For example, the figure shows that the D-S ratio peaks around 12 AM, while the turnover rate peaks around 3 AM. This observation suggests that the D-S ratio is a lagging indicator of E/E behavior, and platform administrators can use this macro indicator to predict changes in user behavior.

{ The consultation of the findings in Section~\ref{sec:ee_user_type} and \ref{sec:ee_user_type} reveals the mechanism behind the stabilization of E/E behavior observed in Figure~\ref{fig:figure2}. Considering that the turnover rate in Section~\ref{sec:ee_user_type} represents a weekly average and the one in Section~\ref{sec:ee_user_type} is the hourly platform average, we observe that users tend to balance exploration and exploitation (turnover rate), even when live-streaming content options are limited depending on the supply of live-streaming content. This suggests that user attributes, such as user age and type, largely shape the balance of E/E behaviors rather than the content itself. Therefore, users are likely selecting live-streaming content and creators based on these personal factors. This result is important, implying that individual parameters play a larger role in shaping E/E behavior rather than platform-wide metrics (e.g., the D-S ratio). However, this analysis reflects the users' average turnover rate and does not imply that users do not change their turnover rates. Section~\ref{sec:res_oshi} reports the analysis of changes in E/E behavior and their impact on user-creator relationships from the perspective of loyalty.}

\subsection{Career Trajectory of Live Streamer}\label{sec:streamer_ranking}

\begin{figure}[ht]
  \centering
  \includegraphics[width=0.5\columnwidth]{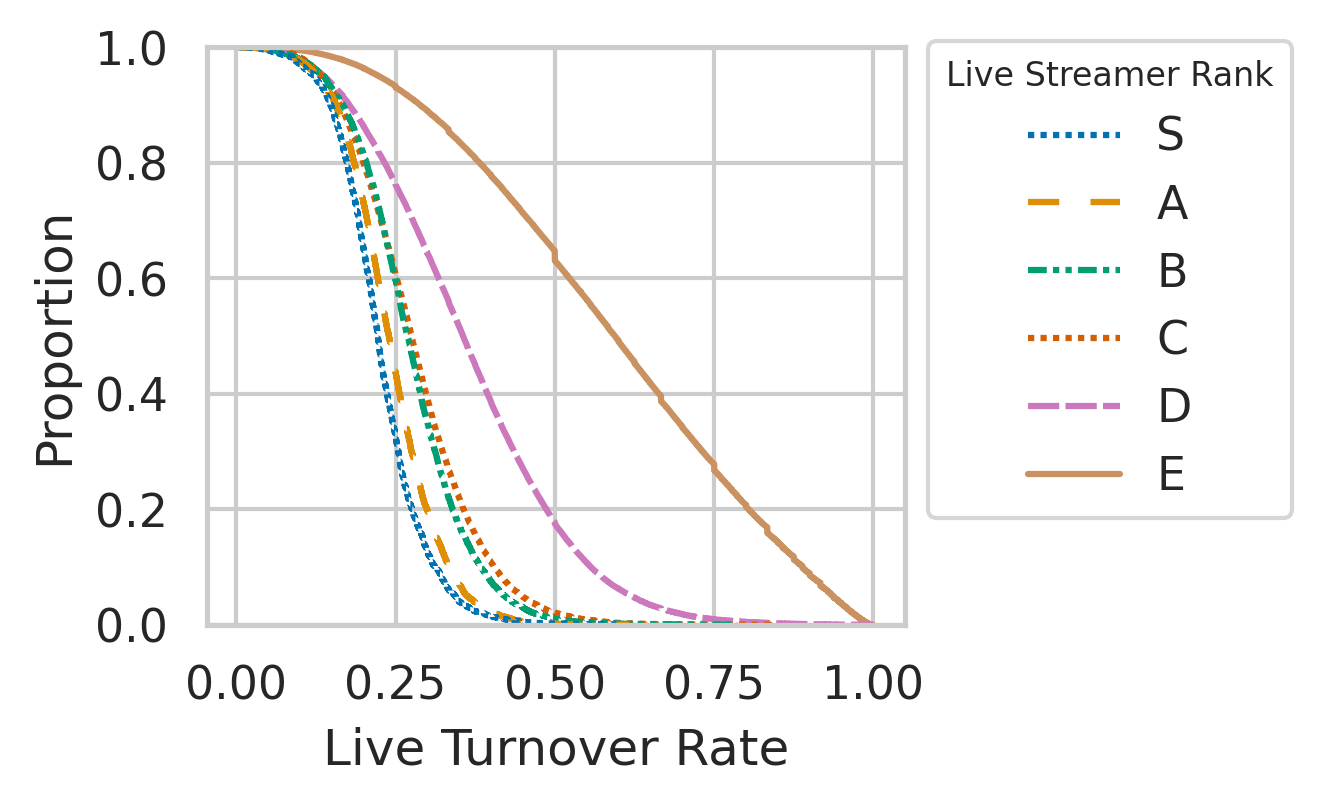}
  \caption{Transition of Live Turnover Rates (CCDF). Complementary cumulative distribution function (CCDF) of the live turnover rate.}
  \label{fig:figure6}
 
\end{figure}

\begin{figure}[htbp]
  \centering
  \includegraphics[width=1\columnwidth]{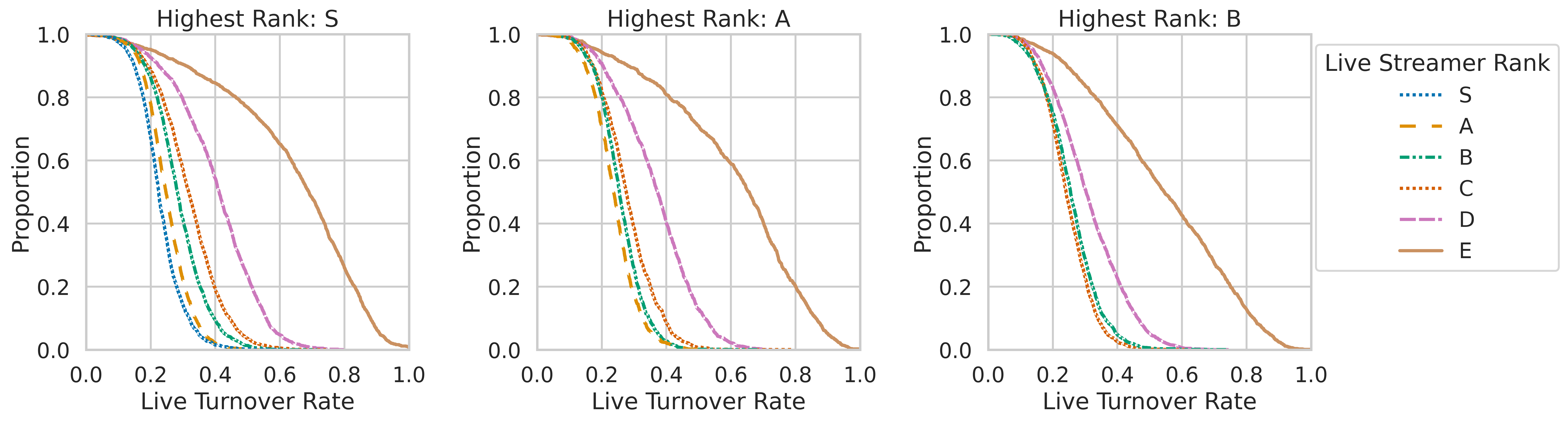}
  \caption{Transition of Live Turnover Rates by Highest Rank  (CCDF). The complementary cumulative distribution function (CCDF) of the live turnover rate per highest rank.}
  \label{fig:figure7}
 
\end{figure}

\begin{figure}[htbp]
  \centering
  \includegraphics[width=0.9\columnwidth]{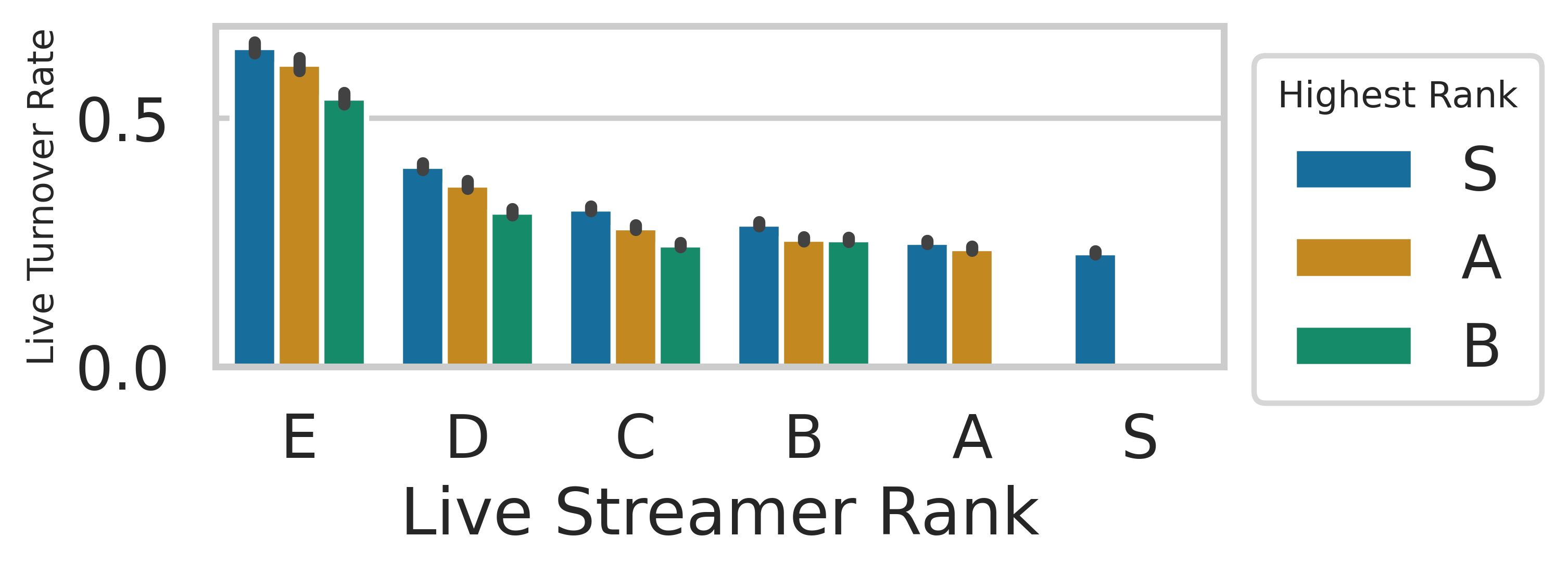}
  \caption{Transition of Live Turnover Rates by Highest Rank}
  \label{fig:figure8}
 
\end{figure}

{
We turn our attention to the live streamers' side and ask, ``Is the degree to which live streamers attract new users associated with the live streamers' performance over their career?'' To address this, we study the career trajectory of live streamers. We calculate the ``live'' turnover rate, which measures the proportion of newcomers in each live session relative to the total number of users who joined that session.

To understand the career trajectory of live streamers and their performance, we track the dynamics of the live turnover rate throughout their careers. We present the results in Figure~\ref{fig:figure6} by plotting the live turnover rate pertaining to rank.
}
The figure illustrates a similar phenomenon where live streamers witness a convergence of live turnover rates (fewer newcomers join their live streaming sessions) as they gain popularity (as indicated by increased live streamer rank). The variability of the live turnover rate also diminishes as the live streamer rank rises as the distribution of the live turnover rate becomes skewed as the live streamer rank increases.

This finding, however, might be influenced by differences in the live streamers' user age, as those achieving higher ranks typically have been active on the platform for longer periods. To handle the potential issue, we employ propensity score matching. We matched users who reached the top rank, S, with those who achieved ranks A and B, based on the number of sessions they streamed. In addition, we focus on live streamers who achieved the top three highest ranks to understand the career trajectory of those with sustained popularity and success.

We plot the live turnover rate using the matched sample in Figures~\ref{fig:figure7} and the observed trend remains consistent. The figure illustrates the evolution of the live turnover rate.
We analyze the live turnover rate corresponding to the rank for each group categorized by their highest achieved rank. For instance, for the group that has achieved the highest rank, S, we calculate the live turnover rate by decomposing it with respect to their current rank. Thus, the live turnover rate for Rank B within the Highest Score S group represents their turnover rate when they are at Rank B. Across the three categories (highest rank S, A, and B), there is a noticeable transition from a dispersed and high distribution to a more skewed and lower distribution as live streamers ascend in the ranking, underscoring the consistency of this pattern in shaping the composition of live streamers' audiences.

The three rank categories exhibit similar trends, but they show different live turnover rates in each rank through their ranking journey and we depict those differences in Figure~\ref{fig:figure8}. The figure shows that the streamers who achieve higher ranks in their career consistently maintain relatively higher live turnover rates. The three groups exhibited different live turnover rates when they were at Ranks E, D, and C. In these bottom three ranks, the higher the highest score, the higher the live turnover rate ($pval < 0.01$). On the other hand, in the top three ranks, we do not find a significant difference in their live turnover rates, with only the group of highest rank S exhibiting a higher turnover rate than others when they were at Rank B ($pval < 0.01$). We also find similar results in the whole sample (i.e., without PS matching) depicted in Figures~\ref{fig:figure10} and \ref{fig:figure11}.

This finding suggests a trade-off between establishing a presence on the platform and the dwindling influx of new users. For mature live streamers (i.e., those with high live streamer rank), their primary audience consists of recurrent users. Live streamers with a low live turnover rate may see many returning live streamers in their sessions. In such scenarios, live streamers tend to spend most of their session time interacting with these recurrent users, potentially not allocating time for new users. In other words, achieving popularity or a solid footing on the platform can, to some extent, may impose some constraints on their content creation. This observation also suggests that achieving a high turnover rate early in a live streamer's career does not necessarily equate to eventual success on the platform, as the highest-ranking live streamers (Rank S) attain higher turnover rates than others on their path to reaching the top rank.

\section{Exploitation Activities and Loyalty: A Case Study with Oshi}\label{sec:res_oshi}

We have analyzed the users' E/E behavior and its importance to the creator side (live streamers). Although we have gained insights into the users' E/E behavior, we have a limited understanding of the link to their relationship with the platform. We dedicate our final analysis to examining users' loyalty to live streamers using turnover rate.

\subsection{Turnover Rate and Oshi}\label{sec:tr_oshi}

\begin{figure}[t]
  \centering
  \includegraphics[width=0.7\columnwidth]{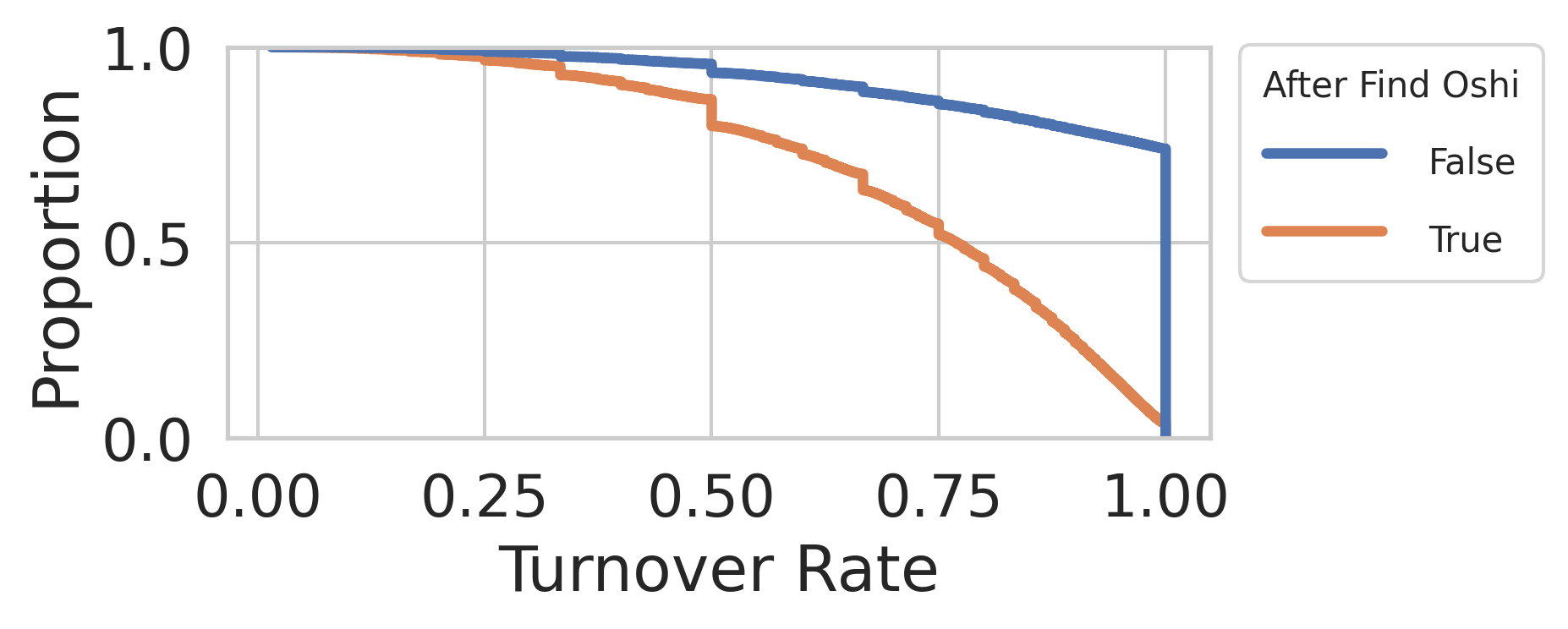}
  \caption{Turnover Change When Meeting Oshi. The distribution of the live turnover rate per user between the month they find the first Oshi and the month before that month by CCDF. The comparison of the mean value of the distribution is $pval < 0.001$.}
  \label{fig:figure9}
 
\end{figure}

We start this subsection by focusing on the first moment users choose an Oshi in their time on the platform and compare the exploration behavior difference between the two consecutive periods before and after they have Oshi. { In this section, we refer to a period as a month, since the set of Oshi is determined monthly.} We plot the distribution of the turnover rate between the two periods in Figure~\ref{fig:figure9}. The figure indicates that the users without Oshi demonstrate high exploration behavior (the mean value of turnover rate is 0.92), and their exploration activity declines when they find the first Oshi (the mean value of turnover rate is 0.72). This clear alternation of turnover rate suggests the connection between the exploration behavior and their loyalty enhancement (Oshi finding). Such users do not only focus on the single live streamer but are engaging in multiple live streams. In fact, we find that the average number of Oshi was about 1.93, which the users select more than two live streamers when they have Oshi.

Next, we examine how users select and change their Oshi. To this end, we conduct a regression analysis to investigate the factors associated with the change of a user's Oshi. To consider the potential confounding factors, we employ a fixed effect regression that removes the differences between users and estimates the average effect. In other words, the fixed effect model in this paper controls time-invariant user-level differences such as preferences, gender, or income, and also time-specific effects that are common to all users at the same time such as platforms' policy change, or seasonal effect, behavioral changes in users during the year-end and New Year's holidays effects. We explained the fixed effect model in Section~\ref{sec:fixed_effect_method}.

We study the relation between E/E behavior and users' loyalty to live streamers by measuring the rate at which they switch their Oshi. To this aim, we calculate the Change Rate \( C_t \) of the composition of Oshi. Let \( M_t \) and \( M_{t+1} \) denote the sets of Oshi live streamers in consecutive months \( t \) and \( t+1 \), respectively, and calculate \( |Q_{t, t+1}|/|M_{t+1}|\) as the Change Rate \( C_t \). Here \( Q_{t, t+1} \) is defined as \( \{m : m \notin M_t, m \in M_{t+1}\} \). The Change Rate \( C_t \) takes the maximum value of 1 when a total replacement of Oshi occurs.

We report the results of the regression analysis in Table~\ref{tab:table2}. We find that the turnover rate is strongly linked to an increase in the Change Rate \( C_t \). The results indicate that approximately a 10.6\% point change in the turnover rate is equivalent to the impact of a 1-hour increase in daily viewing hours on a monthly basis. { We also note the negative coefficient on Viewing Hours per Live Streamer and the positive value of Daily Viewing Hours. This implies that a decrease in the time allocated to each live streamer can lead to a change in their Oshi. The results also indicate that the balance of turnover plays a pivotal role in users' loyalty, not just in consuming the content. For example, given that the number of unique live streamers is log-transformed, a 10\% point change in the turnover rate can contribute to changes in Oshi greater than doubling the number of unique live streamers that a user watched. Note that the turnover rate does not directly represent the users' behavior in their formation of loyalty, rather it depicts their basic consumption behavior (E/E). Nonetheless, our regression model illustrates the E/E behavior embodied by turnover rate suggesting the robust link between the users' E/E behavior and their formation of loyalty to creators (live streamers). } 

This analysis suggests that the turnover rate is intricately linked not just to exploration behavior in general but also to the process through which users develop their loyalty to live-streaming content creators. This identified connection implies that the exploration process reflects a dynamic interplay between user engagement and E/E behavior reflected by the turnover rate. This finding demonstrates that that users' loyalty intensifies not solely by consuming the content of their Oshi but also by actively allocating users' time to live streamers. Consequently, it suggests that understanding these patterns is crucial for live-streaming platforms aiming to enhance user retention and content creator satisfaction.

\begin{table}[htbp]
\centering
\begin{threeparttable}
\caption{Panel OLS Estimation on Oshi Change}
\label{tab:table2}
\begin{tabular}{lrr}
\toprule
Variable & Coefficient &  Std. Err. \\
\midrule

Turnover Rate & ***9.42 & 1.55 \\
\#of Unique Live Streamer (log) & ***1.43 & 0.29 \\
\#of Sessions per Live Streamer & *0.04 & 0.02 \\
\#of Comments per Live Streamer (log) & ***0.83 & 0.23 \\
Viewing Hours per Live Streamer & ***-0.14 & 0.04 \\
Daily Viewing Hours (Base) & ***1.00 & 0.11 \\
\bottomrule
\end{tabular}
\begin{minipage}{\linewidth}
{\footnotesize {\it Note:} Coefficients and standard errors are normalized to the base variable ``Daily Viewing Hours''; *** $pval<0.01$, ** $pval<0.05$, * $pval<0.1$
}
\end{minipage}
 
\end{threeparttable}
\end{table}

\section{Discussion}
This study reveals that E/E behavior on live-streaming platforms follows patterns similar to those observed on on-demand style platforms. The presented research demonstrates that the turnover rate converges to a certain point over time after users join the platform, implying that, even under the real-time interaction constraint, users exploit creators while continuing to explore new content or creators. We compare these results with those from a music listening platform and the analysis indicates that the convergence of the turnover rate occurs more slowly on live-streaming platforms. In other words, users on live-streaming platforms take more time to settle their balance between exploration and exploitation, suggesting that E/E behaviors are influenced by the real-time constraint. Additionally, we detect that external factors, which users cannot directly control, can affect E/E behavior. For instance, supply-demand imbalances introduce fluctuations in the turnover rate. Lastly, E/E behaviors are also influenced by user loyalty to creators (``Oshi'') and the performance of the creators themselves. These findings provide actionable insights for designing live-streaming platforms, particularly those emphasizing the importance of real-time interaction constraints.

By integrating insights from the literature and our findings, we identified opportunities to enhance design strategies for live-streaming platforms. For content creators (live streamers), our analysis first implies that creators should leverage the ``golden period'' to gain new users when they actively explore content at a higher rate. Furthermore, our findings highlight significant changes in E/E behaviors during specific user events that induce large changes in E/E behavior. For instance, we discovered that users, after meeting their first Oshi, decrease their exploration behavior. We also find that loyalty to creators is linked to E/E behavior, but it also shows that users explore new creators even if they have strong loyalty to specific creators (with Oshi). This implies that live streamers face an ambivalent situation: they must not only attract new users but also retain loyal ones. Since the methods for attracting new users often differ from those for engaging existing ones, balancing these strategies is crucial. For example, \cite{wulf2020watching} studied a game streaming platform (Twitch) and argues that streamers tend to have a personal schedule for {\it seeing friends again}. In addition, \cite{hamilton2014streaming} points out that regular streaming (on Twitch) is 
{\it a key role in the formation and growth of stream communities.} Therefore, functions that enable live streamers to monitor users' levels of E/E behavior (turnover) would help them schedule regular streams and host additional unscheduled streams to attract new viewers. In addition, \cite{wohn2020live} points out that direct communication from streamers to viewers is often not feasible because of {\it the mass communication nature of live streaming}. Therefore, some functions—potentially powered by LLMs (Large Language Models)—may assist live streamers in guiding users' E/E behavior.

For platform administrators, our study suggests the platform design should dynamically manage users' E/E behaviors under the real-time interaction constraint. For example, we find that these constraints enlarge the time users take to settle the balance between exploration and exploitation of content compared to on-demand style platforms. In addition, we find that the balance of demand-supply further influences these behaviors. These insights demonstrate that the dynamics of content consumption modeled by E/E behavior in live-streaming platforms are different from on-demand style platforms, implying that simply adapting navigation designs or recommendation algorithms from on-demand consumption models is insufficient for live-streaming platforms. Incorporating explicit E/E logic into recommendation algorithms can also enhance user loyalty and engagement. Additionally, adapting UX designs based on the platform's content supply levels may improve user experience. However, unlike on-demand style platforms, live-streaming platforms have limited options for their content recommendations since the most suitable content to guide users’ E/E behavior may not always be streaming at the moment. 

This study indicates that a design that integrates social interaction with content recommendation may effectively guide and support E/E behavior. Several HCI studies point out that active communication is the most distinct feature of live-streaming platforms, in contrast to the passive engagement typical of on-demand style platforms~\cite{smith2013live, tang2016meerkat}. Therefore, platform designs that manage E/E behavior while maintaining active content consumption and communication require further study. For example, social functions could be leveraged to encourage exploitation behaviors rather than solely relying on content recommendations. \cite{haimson2017makes} emphasize the importance of social networks for user engagement in live-streaming environments. 

The presented study expands the frontier of online user behavior research by revisiting E/E behavior in the novel context of live-streaming platforms and providing rich implications for the human-computer interaction literature. The insights from our analysis will enhance user engagement and foster sustainable ecosystems, benefiting platform operators, content creators, end users, and stakeholders across the online service landscape. Our empirical evidence emphasizes the need for a foundational framework for developing platform designs and algorithms that account for the real-time interaction constraint. We also offer actionable guidance for live-streaming platforms aiming to build effective content recommendation systems and creator support strategies.

\section{Limitations}
While this study offers profound implications for platform management, several limitations remain, as with other academic research. First, our paper focuses on a single platform, and therefore, the generalizability of the findings to other live streaming environments needs further investigation. Future research should encompass multiple platforms and incorporate user surveys to provide a broader understanding of the drivers behind user loyalty and the impact of platform features on user behavior. Additionally, our research does not examine the topics of streaming. Therefore, further analysis considering the topics of discussion and content of the streams that users engage in would offer a more detailed understanding of user behavior, including their E/E behavior. This study highlights the dynamic interplay between exploration and loyalty, emphasizing the need for strategies that support both aspects for sustained platform growth and stability. In addition, our analysis of the turnover rate across user lifespans does not cover instances when users stay very long, such as more than a year. Given recent literature that examines the long-term value of exploration~\cite{suLongTermValue2024}, further research is needed to conduct long-term analyses. Lastly, we also note that the presented analysis is an observational study, meaning that the findings do not necessarily imply causation. Nonetheless, we have employed methods such as propensity scoring and fixed effect models to mitigate potential biases to understand the relationship between turnover rate and metrics related to users or live streamers. Although it is often challenging to detect causal relationships, our robust findings, such as the convergence of the turnover rate, highlight the critical role of balancing exploration and exploitation for real-time interaction platforms.

\section{Conclusion}
Our study reveals that users conduct E/E behavior even on live-streaming platforms under the real-time interaction constraint that they can only consume content being live-streamed at that time. On the one hand, our findings confirm that E/E behavior is general, meaning that the knowledge accumulated on on-demand style platforms can be applicable to live-streaming platforms. On the other, the E/E behavior studied on live-streaming platforms differs in its speed of convergence and is vulnerable to factors that users cannot control. These detected commonalities and differences call for new directions in human-computer interaction research and several important open questions remain unexplored. First, we need to develop an algorithm that induces users' exploration or exploitation by recommending content within the live-streamed content in a real-time environment. Second, we need to examine other connections between E/E behavior and important functionalities. \cite{wohn2020live} discussed the multidimensionality of user-live streamer relationships on live-streaming platforms, but the study does not comprehensively cover all dimensions. For instance, real-time commenting or gifting behavior is tremendously important for users' engagement on live-streaming platforms~\cite{hamilton2014streaming,tang2016meerkat} and it strengthens the engagement~\cite{chen2018drives}. How E/E behavior affects such engagement, however, is still not clear. Filling these gaps in our knowledge would contribute to the design of live-streaming platforms and support content creators in producing effective content, ultimately fostering the development of new cultures and enhancing human communication.

\appendix
 
\renewcommand{\thefigure}{S\arabic{figure}}
\setcounter{figure}{0}
 
\renewcommand{\thetable}{S\arabic{table}}
\setcounter{table}{0}

\section{Supporting information}

\begin{figure}[htbp]
  \centering
  \includegraphics[width=0.7\columnwidth]{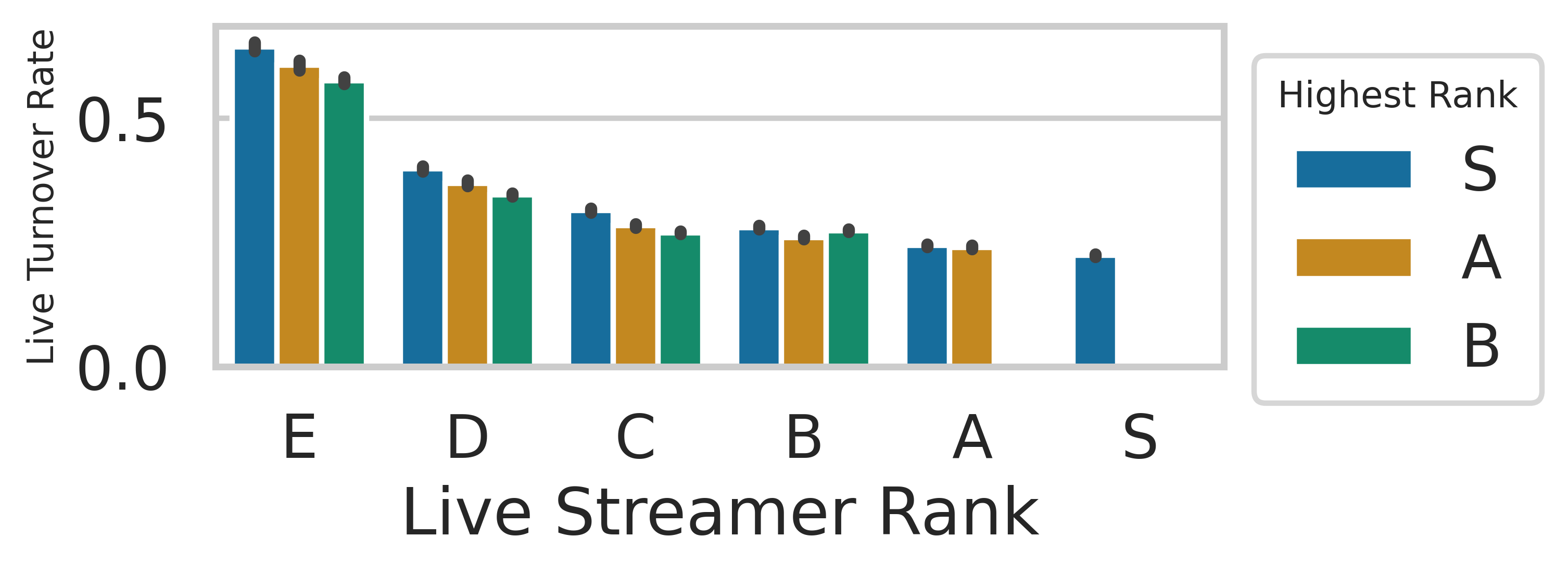}
  \caption{Live Turnover Rate Transition per Final Achieved Live Streamer Rank (whole sample)}
  \label{fig:figure10}
 
\end{figure}

\begin{figure}[htbp]
  \centering
  \includegraphics[width=1\columnwidth]{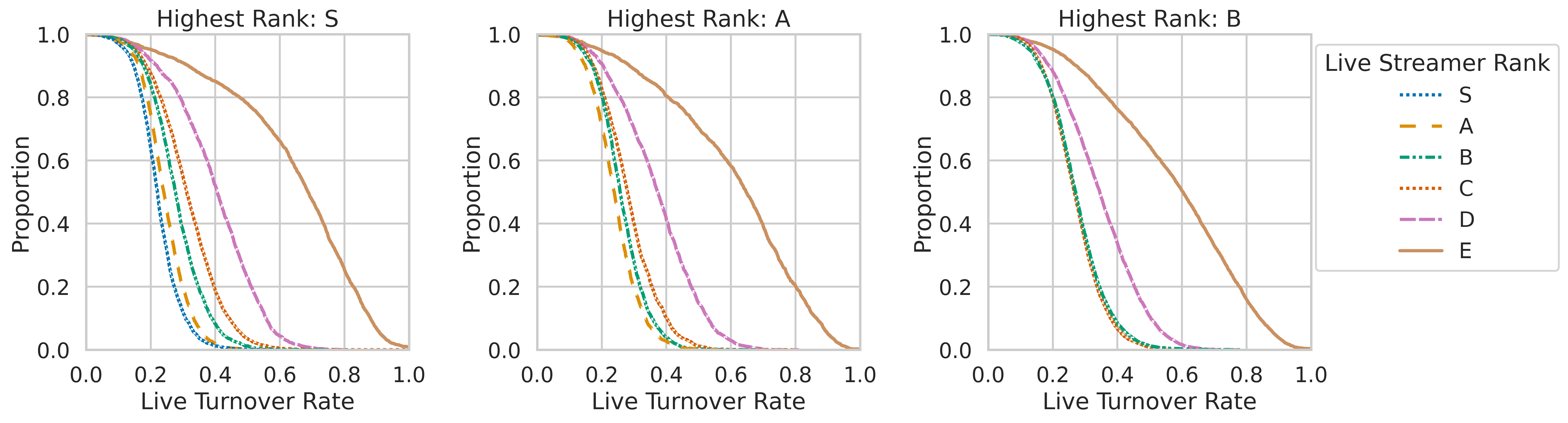}
  \caption{Live Turnover Rate Over the Course of Live Streamer Rank (whole sample)}
  \label{fig:figure11}
 
\end{figure}

\begin{table}[t]
\centering
\begin{tabular}{p{5cm}p{7cm}}
\toprule
Variable & Description \\
\midrule
Turnover Rate & Turnover Rate (monthly average) \\
\#of Unique Live Streamer (log) & \#of unique live streamers that the user watched (log-transformed) \\
\#of Sessions per Live Streamer  & \#of sessions that the user watched divided by the number of unique live streamers \\
\#of Comments per Live Streamer (log)& \#of comments that the user wrote divided by the number of unique live streamers (log-transformed) \\
Viewing Hours per Live Streamer & The amount of time (in hours) that the user watched sessions divided by the number of unique live streamers \\
Daily Viewing Hours (Base) & Daily average of the amount of time that a user spent watching live streaming in that month \\
\bottomrule
\end{tabular}
\caption{Metrics for Live Streaming Platform Analysis}
\label{tab:table3}
\end{table}

\subsection{Turnover Rate in the Day}\label{sec:append_turnover_day}
\begin{figure}[tbp]
  \centering
    \includegraphics[width=1\linewidth]{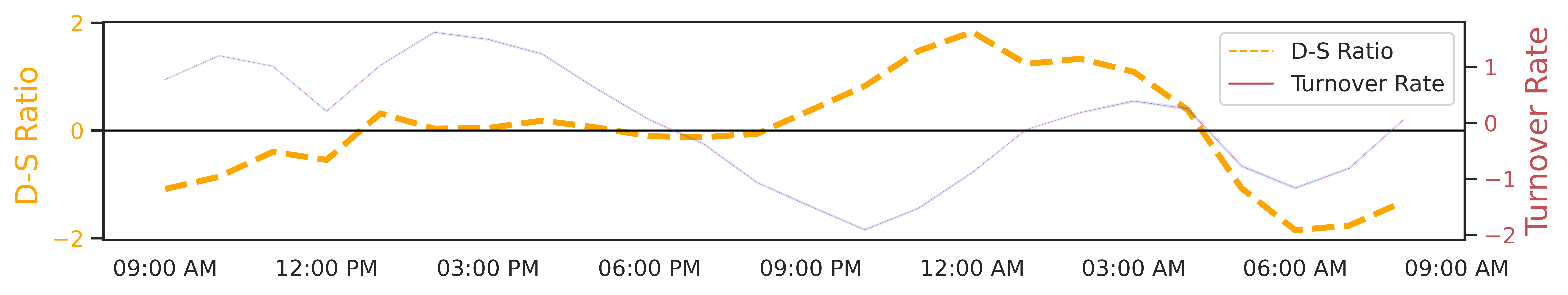}
    \caption{Demand-Supply and Turnover Rate (estimated). The estimated turnover rate (red line) with 95\% intervals (blue shadow).}
    \label{fig:figure12}
\end{figure}

This subsection provides additional information on the turnover rate over the course of the day. We present the turnover rate as estimated by the fixed effect model, as explained in Section~\ref{sec:fixed_effect_method}. Figure~\ref{fig:figure12} calculates the turnover rate while accounting for differences among users through the model. Even in this figure, we still observe fluctuations in the turnover rate throughout the day, similar to those seen in Figure~\ref{fig:figure5}. Moreover, we consistently find a high cross-correlation between the user-live streamer ratio and the turnover rate (0.869 with a lag of -10).

\subsection{Criteria for Oshi}\label{sec:criteria_oshi}

{
On the live streaming platform of this study, the criteria for designating a particular broadcaster as one's ``Oshi'' (favorite streamer) are explicitly described. Users must satisfy one of two sets of requirements within a one-month period to confer Oshi status upon a streamer, and they have to meet the criteria every month to keep their streamers Oshi \footnote{More detail can be found at \url{https://helpfeel.com/pococha-help-us-en/Core_Fan_Requirements-6059740ceea43e00225d0ad5} (Japanese)}.
The first set of criteria to make a live streamer Oshi:
1) An expenditure of 100 yen or more virtual coins (approximately 100 Japanese yen)
, 2)  Watching the live sessions of that streamer for 3 hours or more
, 3) Visiting the live sessions of that streamer on for at least 3 separate occasions
, 4) Posting 10 or more comments on the streamer's live broadcasts The alternative set of criteria to make a live streamer Oshi:
1) An expenditure of 1000 yen or more virtual coins (approximately 1000 Japanese yen)
, 2) Watching the live sessions of that streamer for 0.1 hours or more
, 3) Posting 1 or more comments on the streamer's live broadcasts
}

\section{Acknowledgments}
We pay the utmost attention to the privacy of individual in this study. We did not discuss the results regarding specific users and live streamers, or other individual to maintain the privacy of the individuals studied in this paper. The study in this paper conducted on the data from the beginning of 2020 to the end of 2021, and hence the results in this study may not represent the latest version of the platform. The content of this paper does not represent the current state of the art in the platform, nor does it encapsulate the perspectives and claims of the platform providers who provided the data.

\section*{Funding}
This research was supported by
JSPS KAKENHI Grant Number
JP22K20159, 
JP24K16359. 

\end{document}